\documentclass{article}
\usepackage{ijcai23}

\usepackage{times}
\usepackage{soul}
\usepackage{url}
\usepackage[hidelinks]{hyperref}
\usepackage[utf8]{inputenc}
\usepackage[small]{caption}
\usepackage{graphicx}
\usepackage{amsmath}
\usepackage{amsthm}
\usepackage{booktabs}
\usepackage[switch]{lineno}

\usepackage{algorithm}
\usepackage{algorithmicx}
\usepackage[noend]{algpseudocode}	%

\usepackage{amssymb}
\usepackage{amsthm}
\usepackage{annotate-equations}
\usepackage{xcolor,colortbl}
\usepackage{enumitem}
\usepackage{mathtools}
\usepackage{placeins}     %
\usepackage{subcaption}
\usepackage{relsize}    %
\usepackage{tikz}
\usetikzlibrary{patterns, patterns.meta}
\usepackage{xfrac}
\usepackage{xspace}     %
\usepackage[capitalise]{cleveref}

\newtheorem{example}{Example}

\newtheorem{lemma}{Lemma}
\newtheorem{definition}{Definition}

\DeclarePairedDelimiter\abs{\lvert}{\rvert}%
\newcommand{\graph}{\Gamma}

\newcommand{\assign}{\sigma}                %
\newcommand{\sol}{\sigma}                   %
\newcommand{\support}[1]{sup\left({#1}\right)}

\newcommand{\sensorset}{D}                  %
\newcommand{\maxidset}{k}                   %
\newcommand{\signature}{s}                  %

\newcommand{\sigzero}{S^0}                  %
\newcommand{\sigone}{S^1}                   %

\newcommand{\fullsig}[1]{\left\langle \sigzero_{#1}, \sigone_{#1} \right\rangle}

\newcommand{\assumptions}{\xi}
\newcommand{\auxvarset}{A}                  %
\newcommand{\firevar}{x}                    %
\newcommand{\firevarset}{X}                 %
\newcommand{\sensorvar}{y}                  %
\newcommand{\sensorvarset}{Y}               %
\newcommand{\indvar}{e}                     %
\newcommand{\indvarset}{E}                  %

\newcommand{\indsup}{I}                     %
\newcommand{\groupindsup}{\mathcal{I}}      %
\newcommand{\group}{G}                      %
\newcommand{\groupset}{\mathcal{G}}         %
\newcommand{\unknown}{Q}                    %

\newcommand{\invariant}{C}

\newcommand{\timelimit}{\tau}
\newcommand{\ratio}{r}

\newcommand{\ouralgo}{{\sf\small gismo}}
\newcommand{\ilpmethod}{{\sf\small pbpbs}}

\newcommand{\arjun}{{\sf\small Arjun}}
\newcommand{\bpluse}{{\sf\small B+E}}

\newcommand{\cplex}{{\sf\small CPLEX}}
\newcommand{\cryptominisat}{{\sf\small CryptoMiniSat}}

\newcommand{\github}{{\sf\small GitHub}}

\newcommand{\pblib}{{\sf\small PBLib}}
\newcommand{\python}{{\sf\small Python} $3.5$}

\newcommand{\eg}{\textit{e.g.}}
\newcommand{\ie}{\textit{i.e.}}

\newcommand{\wrt}{\textit{w.r.t.}}
\newcommand{\etal}{\textit{et al.}}
\newcommand{\waylog}{\textit{w.l.o.g.}}

\definecolor{Grey}{gray}{0.85}
\newcolumntype{a}{>{\columncolor{Grey}}c}

\usepackage{algorithm}
\usepackage{algorithmicx}
\usepackage[noend]{algpseudocode}	%
\algnewcommand\algorithmicinput{\textbf{INPUT:}}
\algnewcommand\INPUT{\item[\algorithmicinput]}

\algnewcommand\algorithmicoutput{\textbf{OUTPUT:}}
\algnewcommand\OUTPUT{\item[\algorithmicoutput]}

\algnewcommand{\IIf}[1]{\State\algorithmicif\ #1\ \algorithmicthen}
\algnewcommand{\EndIIf}{\unskip\ \algorithmicend\ \algorithmicif}
\algnewcommand{\IFor}[1]{\State\algorithmicfor\ #1\ \algorithmicdo}
\algnewcommand{\EndIFor}{\unskip\ \algorithmicend\ \algorithmicfor}
\algnewcommand{\IfThenElse}[3]{%
  \State \algorithmicif\ #1\ \algorithmicthen\ #2\ \algorithmicelse\ #3}

\makeatletter
\newcounter{algorithmicH}%
\let\oldalgorithmic\algorithmic
\renewcommand{\algorithmic}{%
  \stepcounter{algorithmicH}%
  \oldalgorithmic}%
\renewcommand{\theHALG@line}{ALG@line.\thealgorithmicH.\arabic{ALG@line}}
\makeatother

\title{Solving the Identifying Code Set Problem with Grouped Independent Support\thanks{\protect Open-source tool, reproducibility info, and extended version of this paper are available at \url{https://github.com/meelgroup/gismo}.}}

\author{
Anna L.D. Latour$^1$\and
Arunabha Sen$^2$\thanks{Work done while Arunabha Sen visited NUS.}\And
Kuldeep S. Meel$^1$
\affiliations
$^1$National University of Singapore, Singapore\\
$^2$Arizona State University, Tempe, AZ, USA
}

\begin{document}

\maketitle

\begin{abstract}
An important problem in network science is finding an optimal placement of sensors in nodes in order to uniquely detect failures in the network.
This problem can be modelled as an {\em identifying code set (ICS)} problem, introduced by Karpovsky \etal{} in 1998.
The ICS problem aims to find a cover of a set $S$, such that the elements in the cover define a {\em unique signature} for each of the elements of $S$, and to minimise the cover's cardinality.
In this work, we study a {\em generalised identifying code set (GICS)} problem, where a unique signature must be found for each {\em subset} of $S$ that has a cardinality of at most $\maxidset$ (instead of just each element of $S$).
The concept of an {\em independent support} of a Boolean formula was introduced by Chakraborty \etal{} in 2014 to speed up propositional model counting, by identifying a subset of variables whose truth assignments uniquely define those of the other variables. 

In this work, we introduce an extended version of independent support, {\em grouped independent support (GIS)}, and show how to reduce the GICS problem to the GIS problem.
We then propose a new solving method for finding a GICS, based on finding a GIS.
We show that the prior state-of-the-art approaches yield integer-linear programming (ILP) models whose sizes grow exponentially with the problem size and $\maxidset$, while our GIS encoding only grows polynomially with the problem size and $\maxidset$.
While the ILP approach can solve the GICS problem on networks of at most $494$ nodes, the GIS-based method can handle networks of up to $21\,363$ nodes; a $\sim 40\times$ improvement.
The GIS-based method shows up to a $520\times$ improvement on the ILP-based method in terms of median solving time.
For the majority of the instances that can be encoded and solved by both methods, the cardinality of the solution returned by the GIS-based method is less than $10\%$ larger than the cardinality of the solution found by the ILP method.
\end{abstract}

\section{Introduction}
\label{sec:introduction}

Imagine that you are in charge of ensuring the fire-safety of a hotel.
Your smoke detectors can sense a fire in the room in which they are placed immediately, and sense a fire in an adjacent room with a time delay.
You realise that this means that you can detect every fire, even if you do not place a detector in every room.
When you tell the hotel manager, they ask you to minimise the number of smoke detectors that you place.
Additionally, they tell you to make sure that, even if as many as five fires break out in different rooms at the same time, you can uniquely identify these multiple rooms based on the set of smoke detectors that detect smoke.
How many detectors do you need, and where do you place them?

The above situation is an example of a {\em sensor placement problem}.
This well-studied problem has applications ranging from satellite deployment~\cite{SGB+19}, to power grid monitoring~\cite{PRP+20}, to identifying criminals~\cite{BS21a} or spreaders of misinformation~\cite{BS21b}, and is typically formulated on graphs.
In the example above, nodes represent the hotel rooms, with edges between adjacent rooms.

Graphs are fundamental tools for modelling the interaction between objects. 
For many real-world computational problems, a node in a graph represents a resource object and an edge between two nodes models the ability for the corresponding objects to communicate. 
Resource objects are often abstractions of critical objects such as satellites, informants in crime networks, or servers. 
The critical nature of these objects necessitates reliable failure detection.
For this, we often rely on sensors, placed strategically on certain nodes.

In this paper, we study a generalised version of the {\em identifying code set (ICS)}~\cite{KCL98} problem.
In our version, a sensor placed in a node detects a failure that occurs in that node immediately, and detects failures in neighbouring nodes with a small time delay.
A {\em generalised identifying code set (GICS)} is a set of nodes in which we must place a sensor such that any set of at most $\maxidset$ simultaneous failures can be {\em uniquely} identified by the placed sensors.
Conceptually, a GICS is a dominating set (\ie{}, a set of nodes such that each node is either in that set or is a neighbour of a node in that set) in an undirected graph, such that each subset of nodes with cardinality at most $\maxidset$ can be uniquely identified by the sensors placed on the nodes this dominating set.

Existing methods for finding and minimising GICSes with one failure at a time, employ an {\em integer-linear programming (ILP)} encoding~\cite{PRP+20,BS21b,BS21a}.
A straightforward generalisation of this ILP formulation to support multiple simultaneous failures scales poorly with network size and the number of simultaneous failures.
This explosion of the model size limits the applicability of ILP-based methods to small networks and support for only one node failure at a time.

The primary contribution of this work is a novel computational technique for solving GICS problems, with a much more compact encoding.
Specifically, we propose the concept of {\em grouped independent support (GIS)} (an extension of {\em independent support}~\cite{CMV14,IMMV16,SM22,YCM22}), and show how we can reduce the problem of finding a GICS to the problem of finding a GIS. We then propose a new algorithm, called {\ouralgo}, to compute a GIS. 

The main benefit of this approach is that the more compact encoding enables us to solve GICS problems on much larger networks than the networks that can be solved by the state of the art.
Indeed, our empirical analysis demonstrates that \ouralgo{} is able to handle networks of up to $21\,363$ nodes, while the ILP-based approach could not handle networks beyond $494$ nodes, thus representing a $\sim 40\times$ improvement in terms of the size of the networks.  
Furthermore, depending on the number of simultaneous failures, the instances that can be encoded by both methods are solved up to $520\times$ faster by the GIS-based approach than by the ILP-based method.
For the majority of those instances, the cardinality of the result returned by \ouralgo{} was at most $10\%$ larger than the cardinality returned by the ILP-based method.

A conceptual contribution is to expand the usefulness of the notion of independent support. 
The computation of independent supports has, to the best of our knowledge, so far only been used as a preprocessing step for model counting and uniform sampling~\cite{CMV14,IMMV16,LLM16,LLM20,YCM22,SM22}. 
We are the first to use the independent support for modelling and solving an NP-hard problem directly.

The remainder of this paper is organised as follows.
We briefly discuss notation and provide relevant definitions in \cref{sec:preliminaries}, where we also provide a motivating example of a GICS problem.
Then, we describe the current state of the art for solving GICS problems in \cref{sec:related-work}.
\Cref{sec:approach} describes GIS, the reduction from the GICS problem to the GIS problem, and \ouralgo{}. 
We present an experimental evaluation of our implementation of \ouralgo{} on a variety of networks in \cref{sec:experiments}, and conclude in \cref{sec:conclusion}.

\section{Preliminaries}
\label{sec:preliminaries}

We briefly introduce our notation, recall relevant concepts, and define the {\em generalised identifying code set (GICS)}.

\subsection{Definitions and Notation}
\label{subsec:definitions-and-notation}

\paragraph{Graphs.}
We consider an undirected, loop-free graph $\graph=(V,E)$ on nodes $V$ and edges $E$. 
We denote nodes with lower case letters $u,v,w \in V$. 
The distance between two nodes $u$ and $v$ is the number of edges on the shortest path between them, and is denoted by $d(u,v)$.
If $d(u,v) = 1$, we call the nodes $u$ and $v$ {\em direct neighbours} of each other.
The neighbourhood function $N_d(v)$ returns the set of nodes that are at a distance $d$ from node $v$.
We define the \textit{closed $d$-neighbourhood} of a node $v$ as $N_d^+(v) = N_d(v) \cup \{v\}$.
For a set of nodes $U$, we define the neighbourhood function $N_d(U) = \bigcup_{u \in U} N_d(u)$, and define the closed neighbourhood of $U$, $N_d^+(U)$, analogously.

\paragraph{Boolean satisfiability.}
We denote a set of Boolean variables with the capital letter $X$ and denote individual Boolean variables with lowercase letters $x, y, z \in X$.
We denote truth values with $1$ ({\em true}) and $0$ ({\em false}).
A literal $l$ is a variable (\eg{}, $x$) or its negation (\eg{}, $\neg x$).
A disjunction of literals is called a {\em clause}.
We say that a formula $F$ is in {\em conjunctive normal form} (CNF) if it is a conjunction of clauses.
A {\em full assignment} $\assign: X \mapsto \{0,1\}$ assigns a truth value to each variable in $X$.
We use $\assign(x)$ to denote the truth value that $\assign$ assigns to variable $x$.
Given a subset $Y \subseteq X$, $\assign_{\downarrow Y}: Y \mapsto \{0,1\}$ denotes the assignment {\em projected} onto $Y$, thus specifying the truth values that the variables in $Y$ get under $\assign$.
Given a Boolean formula $F(X)$, we call an assignment $\sol$ a {\em solution} or {\em model} of $F(X)$ if $F(\{x \mapsto \sol(x) \mid x \in X  \}) \models 1$.
We denote the set of all models of $F$ with $Sol(F)$.
Similarly, we denote the set of all models of $F(X)$ projected on the subset $Y \subseteq X$ as $Sol_{\downarrow Y}(F)$.
We call the variables $X$ that appear in $F$ the {\em support} of $F$. 
If a Boolean formula has at least one solution, we say that it is {\em satisfiable}.
Otherwise, we call it {\em unsatisfiable}.

\paragraph{Minimality.}
Let $T$ be a set of items, and let $S \subseteq T$ be a subset.
Given a set $\mathcal{C}$ of constraints on sets, we call $S$ {\em set-minimal} \wrt{} $\mathcal{C}$ if $S$ satisfies all constraints in $\mathcal{C}$ and there exists no proper subset of $S$ that also satisfies all those constraints.
We call $S$ a {\em cardinality-minimal} set if $S$ is minimal, and there exists no $S^\prime \subseteq T$ that is also minimal, but whose cardinality is strictly smaller than that of $S$.

\paragraph{Support of a set.}
We use calligraphic uppercase symbols to denote sets of sets of variables. We define the {\em support} of a set of sets of variables $\mathcal{S}$ as follows: $\support{\mathcal{S}} := \bigcup_{S_i \in \mathcal{S}} S_i$.

\paragraph{Signatures.}
Given an undirected, loop-free graph $\graph := (V, E)$ with nodes $V$ and edges $E$, and given a subset of nodes $\sensorset \subseteq V$.
We define the {\em signature} of $U \subseteq V$ as the following tuple: $\signature_U := \fullsig{U}$, where $\sigzero_U := U \cap \sensorset$ and $\sigone_U := N_1^+(U) \cap \sensorset$.

\paragraph{Generalised Identifying Code Set (GICS).}
Given a graph $\graph:=(V,E)$, a positive integer $\maxidset \leq \abs{V}$ and $\sensorset \subseteq V$.
We call $\sensorset$ a {\em generalised identifying code set (GICS)} of $\graph$ and $\maxidset$ if, for all $U, W \subseteq V$ with $\abs{U} \leq \maxidset$, $\abs{W} \leq \maxidset$ and $U \neq W$, it holds that $\signature_U \neq \signature_W$.
Hence, if $\sensorset$ is a GICS of $\graph$ and $\maxidset$, then the signatures of all subsets of $V$ with cardinality at most $\maxidset$ are unique.
We call $\maxidset$ the {\em maximum identifiable set size}.

\paragraph{The GICS problem.}
Given a $\graph:=(V,E)$ and $\maxidset$, the GICS problem asks to find a $\sensorset \subseteq V$ such that $\sensorset$ is a GICS of $\graph$ and $\maxidset$, and $\abs{\sensorset}$ is minimised.

\paragraph{Independent Support.}
Given a Boolean formula $F(X)$ and a set $\indsup \subseteq X$, we call $\indsup$ an {\em independent support}~\cite{CMV14} of $F$ iff, for two solutions $\sol_1$ and $\sol_2$, the following holds: $\left(\sol_{1\downarrow\indsup} = \sol_{2\downarrow\indsup} \right) \Rightarrow \left(\sol_{1\downarrow X} = \sol_{2\downarrow X} \right)$.

The concept of independent support was introduced in 2014~\cite{CFM+14}, born from the observation that the truth values assigned to variables in solutions to a formula, can often be defined by the truth values of other variables.
Hence, this property is referred to in the literature as {\em definability}~\cite{LLM16,SM22}.
Tools for computing minimal independent supports include \arjun{}~\cite{SM22} and \bpluse{}~\cite{LLM16,LLM20}.

Until now, independent supports have only been computed as a preprocessing step for counting and sampling~\cite{CFM+14,LLM16,LLM20,CMV14,IMMV16,SM22,YCM22}.
In \cref{sec:approach}, we present a generalisation of the independent support of a Boolean formula, and show how we can use that to find solutions to the GICS problem.
To the best of our knowledge, we are the first to lift computing independent supports out of the preprocessing domain, turning it into a tool for modelling and solving NP-hard problems directly.

\paragraph{Padoa's Theorem.}
Let $F(Z, \auxvarset)$ be a Boolean formula on Boolean variables $Z \cup \auxvarset$, with $Z \cap \auxvarset = \varnothing$.
We can use Padoa's theorem~\cite{Pad1901} to check if a variable $z \in Z$ is defined by the other variables in $Z$.
Let $\hat{Z}$ be a fresh set of variables, such that $\hat{Z} := \{\hat{z}_i \mid z_i \in Z\}$, and let $F\left(Z\mapsto \hat{Z}, \auxvarset\right)$ be the formula in which every $z_i \in Z$ is replaced by its corresponding $\hat{z}_i \in \hat{Z}$.
We assume \waylog{} that $Z := \{z_1, \ldots, z_{m}\}$, with $m = \abs{Z}$.
For $1 \leq i \leq m$, Padoa's theorem now defines the following formula:
\begin{equation}
    \begin{aligned}
        \psi\left(Z, \auxvarset, \hat{Z}, i\right) :=\: &F(Z, \auxvarset) \land F\left(Z \mapsto \hat{Z}, \auxvarset\right) \:\land \\
        &\bigwedge_{j = 1;j\neq i}^{m} \left(z_j \leftrightarrow \hat{z}_j\right) \land z_i \land \neg \hat{z}_i.
    \end{aligned}
    \label{eq:padoa}
\end{equation}
Intuitively, this formula asks if there exist at least two solutions to $F(Z, \auxvarset)$, $\sol_1$ and $\sol_2$, such that $\sol_{1\downarrow Z}$ and $\sol_{2 \downarrow Z}$ differ only in their value for $z_i$.
If yes, then \cref{eq:padoa} is satisfiable.
If no, then \cref{eq:padoa} is unsatisfiable.

\subsection{Motivating Example}
\label{subsec:motivating-example}

We model the sensor placement example from \cref{sec:introduction} as follows.
First, we model the hotel as a graph $\graph := (V,E)$, where the nodes $V$ represent rooms and two nodes $u,v$ are connected by an edge $(u,v)$ if the corresponding rooms are adjacent.
Smoke detectors have a green light if they do not detect smoke, and have a red light if they do.
All smoke detectors have a green light at $t_0 - 1$.
We assume that at time $t_0$ a fire can break out in at most $\maxidset$ different rooms ($\maxidset = 5$ in the example in \cref{sec:introduction}), and that after $t_0$, no more fires break out.
If there is a smoke detector placed in room $v$, and a fire breaks out in room $v$ at time $t_0$, the smoke detector in room $v$ detects the smoke at $t_0$, whereupon its detection light turns from green to red immediately, and remains red.
A smoke detector placed in room $u \in N_1(v)$ detects the smoke from the fire in room $v$ at $t_1 = t_0 + 1$.
If its light was not yet red at time $t_0$, the light of the sensor in room $u$ turns from green to red at time $t_1$.
Hence, at time $t_1$ a sensor placed in room $v \in V$ is red iff there is a fire in at least one room in $N_1^+(v)$.

For a set of rooms $U \subseteq V$, we now have $\signature_U = \fullsig{U}$, where $\sigzero_U$ represents the set of detectors whose lights turn red at $t_0$ if fires break out in all rooms in $U$ at $t_0$, while $\sigone_U$ represents the set of detectors whose lights are red at $t_1 = t_0 + 1$.
The GICS problem asks in which set of nodes $\sensorset \subseteq V$ to place a smoke detector, such that $\sensorset$ is a GICS of $\graph$ and $\maxidset$, and $\abs{\sensorset}$ is minimised.

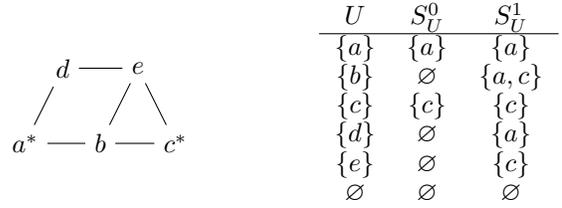
\begin{figure}[t]
    \centering
    \begin{tikzpicture}
        \node (a) at (0, 0) {$a^*$};
        \node (b) at (1, 0) {$b$};
        \node (c) at (2, 0) {$c^*$};
        \node (d) at (0.5, 1) {$d$};
        \node (e) at (1.5, 1) {$e$};
        
        \path 
        (a) edge (b)
            edge (d)
        (b) edge (c)
            edge (e)
        (c) edge (e)
        (d) edge (e);
        
        \node (table) at (5.5, 0.5) {
            \begin{tabular}{ccc}
                $U$ & $\sigzero_U$ & $\sigone_U$ \\ \hline
                $\{a\}$  & $\{a\}$  & $\{a\}$ \\
                $\{b\}$  & $\varnothing$  & $\{a,c\}$ \\
                $\{c\}$  & $\{c\}$  & $\{c\}$ \\
                $\{d\}$  & $\varnothing$  & $\{a\}$ \\
                $\{e\}$  & $\varnothing$  & $\{c\}$ \\
                $\varnothing$ & $\varnothing$  & $\varnothing$
            \end{tabular}
        };
    \end{tikzpicture}
    \caption{A graph and a GICS $\sensorset := \{a,c\}$ for $\maxidset = 1$, indicated with $*$. The table shows the signatures for all $U \subseteq V$ with $\abs{U} \leq \maxidset$.}
    \label{fig:example-spp-1}
\end{figure}

\begin{example}
\label{ex:small-network}
\Cref{fig:example-spp-1} shows an example of five rooms, where we have chosen $\maxidset=1$.
We show a GICS for this example that places a sensor in rooms $a$ and $c$, \ie{}, $\sensorset :=\{a,c\}$.
The table shows the signature for each subset $U \subseteq V$ with $\abs{U} \leq 1$.
Note that each signature is unique, each non-empty subset has a non-empty signature, and that neither $a$ nor $c$ can be removed from $\sensorset$ without destroying these two properties.
This particular GICS has cardinality $2$, which is the smallest possible cardinality for this network with $\maxidset = 1$.
\end{example}

\section{Related Work}
\label{sec:related-work}

Several methods have been proposed for solving a variant of the identifying code set problem that only considers a maximum identifiable set size of $\maxidset=1$, and only requires the $\sigone_U$s to be unique.
A common approach~\cite{SGB+19,PRP+20,BS21b,BS21a} models the problem as an {\em integer-linear program (ILP)}, to be solved with a {\em mixed-integer programming (MIP)} solver.
We adapted the method from~\cite{PRP+20,BS21a} such that it can model the unique identification of $k > 1$ simultaneous events.
The number of linear constraints in this encoding grows as $O\left(\binom{\abs{V}}{\maxidset}^2\right)$, which is prohibitively large for all but the smallest of networks, especially if $\maxidset > 1$.
We refer the reader to the extended version of this paper for the details on this ILP encoding and its size.

\section{Approach}
\label{sec:approach}

In this section, we discuss our novel approach to solving the GICS problem, which uses an encoding whose size {\em does not} explode, but rather grows polynomially with the problem size and $\maxidset$.
We first introduce the {\em grouped independent support (GIS)}, an extension of {\em independent support}~\cite{CFM+14,IMMV16}, then show how we can reduce finding a GICS to finding a GIS, and finally propose an algorithm for finding a GIS of minimised cardinality: \ouralgo{}.

\subsection{Grouped Independent Support (GIS)}

We define {\em grouped independent support (GIS)} as follows:
\begin{definition}
\label{def:gis}
Given a formula $F(Z,\auxvarset)$, with $Z \cap \auxvarset = \varnothing$ and a partitioning $\groupset$ of $Z$ into non-empty sets, such that $\support{\groupset} = Z$.
The subset $\groupindsup \subseteq \groupset$ is a {\em grouped independent support} of $\langle F(Z, \auxvarset), \groupset \rangle$ if the following holds:
\begin{equation}
    \begin{aligned}
        &\forall \sol_1, \sol_2 \in Sol(F)\\
        &\left(\left(\sol_{1 \downarrow \support{\groupindsup}} = \sol_{2 \downarrow \support{\groupindsup}}\right) \leftrightarrow \left(\sol_{1 \downarrow Z} = \sol_{2 \downarrow Z}\right)\right).
    \end{aligned}
    \label{eq:indep-condition}
\end{equation}
\end{definition}
Intuitively, this means that if all solutions projected on to $Z = \support{\groupset}$ are unique, then all solutions projected onto $\support{\groupindsup} \subseteq \support{\groupset}$ are unique, and vice versa.
The `$\rightarrow$' in \cref{eq:indep-condition} means that, for all solutions to $F(Z, \auxvarset)$, the truth values of the variables in $\support{\groupset} \setminus \support{\groupindsup}$ are defined by the truth values of the variables in $\support{\groupindsup}$.
Note that GIS is a generalisation of independent support, since finding an independent support corresponds to finding a GIS where all the groups have cardinality $1$.

Observe that the problem of checking whether a given set $\mathcal{I}$ is a grouped independent support is in co-NP.
In contrast, checking whether an assignment satisfies ILP constraints is in polynomial time. 
Therefore, a priori, it is natural to wonder if it is worth reducing GICS to a (potentially) computationally harder problem. 
We pursue such a reduction in the hope that the reduction may come at the gain of smaller problem encodings. 
In the remainder of this section, we show that such a gain is indeed possible. 

\subsection{A Reduction from GICS to GIS}
\label{subsec:reduction}

We now present a reduction from finding a GICS to finding a GIS, using the example problem from \cref{subsec:motivating-example}.
Let $\firevarset := \{\firevar_v \mid v \in V\}$ and $\sensorvarset := \{\sensorvar_v \mid v \in V\}$ be sets of Boolean variables such that $\firevar_v = 1$ and $\sensorvar_v = 1$ iff a sensor placed in room $v$ has a red light at time $t_0$ and $t_1$, respectively.
We capture this in the following Boolean formula:
\begin{equation}
    F_{detection} := \bigwedge_{v \in V} \left(\sensorvar_v \leftrightarrow \bigvee_{u \in N_1^+(v)} \firevar_u\right).
    \label{eq:detection}
\end{equation}
Additionally, we must require that at most $\maxidset$ fires break out at the same time, which we do with the following formula (recall that if a fire breaks out at time $t_0$ in room $v$, the light of a sensor placed in room $v$ turns red at $t_0$):
\begin{equation}
    F_{card,\maxidset} := \sum_{v \in V} \firevar_v \leq \maxidset.
    \label{eq:cardinality}
\end{equation}
Converting these constraints to CNF and conjoining them, we obtain the following formula in CNF:
\begin{equation}
    F_\maxidset(\firevarset \cup \sensorvarset, \auxvarset) = F_{detection} \wedge F_{card,\maxidset},
    \label{eq:cnf}
\end{equation}
where $\auxvarset$ is a (possibly empty) set of auxiliary variables needed for the CNF encoding of the cardinality constraint. 
Additionally, we define one group for each node in the network: $\groupset := \{\group_v := \{\firevar_v, \sensorvar_v\} \mid v \in V \}$.

Now, we can find a GICS by encoding the problem into CNF according to \cref{eq:cnf}, finding a GIS, and then extracting the sensor set as: $\sensorset := \{ v \mid \group_v \in \groupindsup\}$.

\begin{lemma} 
    \label{lem:from-GIS-to-GICS}
    Given a loop-free, undirected network $\graph := \left(V, E\right)$ on nodes $V$ and edges $E$, a maximum identifiable set size $0 < \maxidset \leq \abs{V}$, and given a GIS of \cref{eq:cnf} $\groupindsup \subseteq \groupset$ with groups $\groupset := \{\group_v := \{x_v, y_v\}\}$.
    The set $\sensorset := \{v \in V \mid \group_v \in \groupindsup\}$ is a GICS of $\graph$.
\end{lemma}
We prove this lemma in the extended version, by proving that there is a bijective relationship between the elements of the set of signatures of all $U \subseteq V$ with $\abs{U} \leq \maxidset$ and the set of projected solutions $Sol_{\downarrow \support{\groupindsup}}\left(F_\maxidset\right)$.
Intuitively, we show that the highlighted columns in \cref{tab:truth-table} encode the signatures in the table in \cref{fig:example-spp-1}, and vice versa.
Each solution to $F_\maxidset$ in \cref{eq:cnf} corresponds to selecting a set $U \subseteq V$ with $\abs{U} \leq \maxidset$ as the set of nodes with failures.
By \cref{def:gis}, $\sol_{U\downarrow \support{\groupindsup}} \neq \sol_{W\downarrow \support{\groupindsup}}$ for sets $U,W \in V$ with $\abs{U}, \abs{W} \leq \maxidset$. 
Therefore, a GIS $\groupindsup$ guarantees unique signatures for all such $U,W$. 
Hence, the uniqueness requirement is implicitly required by the semantics of GIS, and we do not need to encode it explicitly.

We can prove the following lemma by simple analysis of \cref{eq:cnf} and techniques for encoding cardinality constraints into CNF~\cite{sin05,PS15}, and refer the reader to the extended version of this paper for that proof:

\begin{lemma}
    \label{lem:num-clauses}
    $F_\maxidset(\firevarset \cup \sensorvarset, \auxvarset)$ has $O\left( \maxidset \cdot \abs{V} + \abs{E} \right)$ clauses.
\end{lemma}

The above lemma highlights the potential exponential gains in encoding from GIS-based approach.
While the ILP-based approach would lead to encodings with $O\left(\binom{\abs{V}}{\maxidset}^2\right)$ constraints, our GIS-based approach requires only $O\left(\maxidset \cdot \abs{V}  + \abs{E} \right)$ clauses. 

\subsection{\protect Finding a GIS with {\sf gismo}}
\label{subsec:finding-a-gis}

\Cref{alg:find-minimal-gis} shows our algorithm for finding a GIS.
On a high level, the algorithm iterates over all groups of variables and uses Padoa's theorem to determine if at least one of the variables in each variable group is {\em not} defined by other variables outside the group.
If this is the case, the group must be part of a GIS.
We now describe \ouralgo{} in more detail.

\begin{algorithm}[tb]
\caption{The \ouralgo{} algorithm.}
\label{alg:find-minimal-gis}
 \textbf{Input:} Formula $F(Z, \auxvarset)$ with partitioning of $Z$ into $\groupset$, a time limit $\timelimit$.\\
 \textbf{Output:} GIS $\groupindsup \subseteq \groupset$.\\[-2mm]

\begin{algorithmic}[1] %
\State $E \gets \{e_i \mid z_i \in Z\}$ \label{line:init-indicatorvars}
\State Initialise $\phi\left(Z, \auxvarset, \hat{Z}\right)$ \label{line:init-phi} \Comment{\cref{eq:phi}}
\State $\unknown \gets \support{\groupset}$ \label{line:init-unknown}
\State $\groupindsup \gets \varnothing$ 
\For{$\group \in \groupset$} \label{line:start-outer-for} 
    \State $\unknown \gets \unknown \setminus \group$ \label{line:update-Q} 
    \State $\invariant \gets \unknown \cup \support{\groupindsup}$ 
    \State $\assumptions \gets \bigwedge_{z_i \in \invariant} \indvar_i$  
    \For{$z \in \group$} \label{line:start-inner-for} 
        \State $\psi \gets \phi \land \xi \land z \land \neg \hat{z}$ \Comment{note similarity to \cref{eq:padoa}} \label{line:init-psi}
        \State $sat \gets \textsc{CheckSAT}(\psi, \timelimit)$
        \If {$sat$} \label{line:is-independent}
            \State $\groupindsup \gets \groupindsup \cup \{\group\}$ \label{line:update-groupindsup} 
            \State break \label{line:break}
        \EndIf
    \EndFor
\EndFor
\State \Return $\groupindsup$
\end{algorithmic}
\end{algorithm}

Recall Padoa's theorem from \cref{subsec:definitions-and-notation}.
By choosing $Z := \support{\groupset} = \firevarset \cup \sensorvarset$, we can define $\psi$ for \cref{eq:cnf}.
If, for an $1 \leq i \leq m$, $\psi\left(Z, \auxvarset, \hat{Z}, z_i\right)$ is unsatisfiable, then we know the following: if a partial assignment $\sol_{\downarrow Z \setminus \{z_i\}}$ can be extended to $\sol_{\downarrow Z}$, then there is only one possible value that $z_i$ can take in $\sol_{\downarrow Z}$ such that $\sol$ is a model of \cref{eq:cnf}.
Hence, if $\psi$ is unsatisfiable, then for each $\sol$, the truth value of $z_i$ is defined by the truth values of the variables $Z \setminus \{z_i\}$.

In \cref{alg:find-minimal-gis}, we use Padoa's theorem as follows.
We introduce a fresh set of indicator variables $\indvarset = \{\indvar_i \mid z_i \in Z\}$ (\cref{line:init-indicatorvars}), and define the following formula (\cref{line:init-phi}):
\begin{equation}
    \begin{aligned}
        &\phi\left(Z, \auxvarset, \hat{Z}\right) := \\ 
        &\quad F(Z, \auxvarset) \land F\left(Z \mapsto \hat{Z}, \auxvarset\right) \land \bigwedge_{\substack{j = 1}}^{m} \indvar_j \rightarrow \left(z_j \leftrightarrow \hat{z}_j\right).
    \end{aligned}
    \label{eq:phi}
\end{equation}
In \cref{line:init-unknown}, we introduce $\unknown$, the set of candidate variables that could be in the support of the GIS $\groupindsup$ that is returned by \ouralgo{}. 
We initialise $\groupindsup$ with $\varnothing$.

The `for'-loop that starts at \cref{line:start-outer-for} in \cref{alg:find-minimal-gis} iterates over the groups in partition $\groupset$.
In each iteration, we define the set $\invariant$, which contains the variables for which we want to check if they define the variables in the group $\group$ that is considered in that iteration.
By design, the set $\invariant \cup \group$ is an independent support of $F(Z, \auxvarset)$.
In the `for'-loop that starts at \cref{line:start-inner-for}, we test for each variable $z \in \group$ if that variable is defined by the variables in $\invariant$, and thus if $\invariant$ is an independent support.
If $z$ is {\em not} defined by the variables in $\invariant$ (and hence $\psi$ is satisfiable), we know that, given the current $\invariant$, $z$ is needed to define all solutions, and thus that $\invariant$ is {\em not} an independent support of $F(Z, \auxvarset)$.
Hence, we add $z$'s entire group to the GIS $\groupindsup$ (in \cref{line:update-groupindsup}).
If all variables in $\group$ are defined by the variables in $\invariant$, then $\invariant$ is an independent support and none of the variables in $\group$ are needed to define all solutions, so $\group$ is not added to $\groupindsup$, and not considered again.

At the start of each iteration of the outer `for'-loop, $\unknown \cup \support{\groupindsup}$ is a set of variables that define the variables in $Z \setminus \left(\unknown \cup \support{\groupindsup}\right)$.
During the execution of the algorithm, more and more groups of variables are removed from $\unknown$, and some groups are added to $\indsup$, if that is deemed necessary for $\unknown \cup \support{\groupindsup}$ to still define the variables in $Z \setminus \left(\unknown \cup \support{\groupindsup}\right)$.
At the end of the algorithm, $\unknown$ is empty, and hence all groups in $\groupindsup$ contain variables that are necessary for defining the variables in $Z \setminus \support{\groupindsup}$.
Hence, the $\groupindsup$ returned by the algorithm is a GIS for $\langle F(Z, \auxvarset), \groupset\rangle$.
Recall that $\sensorset$ is a GICS, and that in our reduction, each group corresponds to a node. 
Therefore, intuitively, \ouralgo{} starts with $\sensorset = V$, and then removes nodes from $\sensorset$ until no nodes can be removed without removing the GICS-ness of $\sensorset$.

The time limit $\timelimit$ in \cref{line:is-independent} is given in a maximum number of conflicts that the SAT solver may encounter before giving up.
If the SAT solver reaches $\timelimit$ before it determines the (un)satisfiability of $\psi$, then $\psi$ is treated as satisfiable.
Hence, in practice it may happen that $\group$ is defined by the variables in $\invariant$, but is nevertheless added to $\groupindsup$.

We refer the reader to the extended version of this paper for the proof of the following lemma:
\begin{lemma}
    \label{lem:algo-gis}
    Given an input formula $F(Z, \auxvarset)$ with group partitioning $\groupset$ such that $\support{\groupset} = Z$, \cref{alg:find-minimal-gis} returns a GIS $\groupindsup$ of $\langle F(Z, \auxvarset), \groupset \rangle$.
\end{lemma}

If the call to {\sc CheckSAT($\psi$, $\tau$)} never times out, \ouralgo{} returns a set-minimal GIS of the input formula and partition.
The cardinality of that GIS is potentially larger than the cardinality-minimal solution that is guaranteed by the ILP encoding proposed by \cite{PRP+20}.

Note the similarity of \ouralgo{} to the algorithm for high-level minimal unsatisfiable core extraction presented in \cite{Nad10}.
Indeed, finding a set-minimal independent support can be reduced to finding a group-oriented (or high-level) minimal unsatisfiable subset~\cite{IMMV16}. 

We illustrate \ouralgo{} with an example, based on the problem in \cref{ex:small-network}.
To aid our discussion, we provide a truth table containing all solutions to \cref{eq:cnf} for the problem in \cref{ex:small-network}, in \cref{tab:truth-table}.

\begin{table}[t]
    \centering
    \small
    \begin{tabular}{c|acacc|acacc}
        ~ &\multicolumn{5}{c|}{$\firevarset$, or $\sigzero_U$}    & \multicolumn{5}{c}{$\sensorvarset$, or $\sigone_U$} \\
        $U$ & $\firevar_a$   & $\firevar_b$ & $\firevar_c$ & $\firevar_d$ & $\firevar_e$ &   $\sensorvar_a$   & $\sensorvar_b$ & $\sensorvar_c$ & $\sensorvar_d$ & $\sensorvar_e$ \\ \hline
        $\{a\}$ &   $1$ & $0$    & $0$   & $0$   & $0$  & $1$    & $1$   & $0$   & $1$   & $0$ \\
        $\{b\}$ &   $0$ & $1$    & $0$   & $0$   & $0$  & $1$    & $1$   & $1$   & $0$   & $1$ \\
        $\{c\}$ &   $0$ & $0$    & $1$   & $0$   & $0$  & $0$    & $1$   & $1$   & $0$   & $1$ \\
        $\{d\}$ &   $0$ & $0$    & $0$   & $1$   & $0$  & $1$    & $0$   & $0$   & $1$   & $1$ \\
        $\{e\}$ &   $0$ & $0$    & $0$   & $0$   & $1$  & $0$    & $1$   & $1$   & $1$   & $1$ \\
        $\varnothing$ &   $0$ & $0$    & $0$   & $0$   & $0$  & $0$    & $0$   & $0$   & $0$   & $0$
    \end{tabular}
    \caption{The rows of the truth table of \cref{eq:cnf} that correspond to models of \cref{eq:cnf}, for the small graph in \cref{fig:example-spp-1}. The grey columns highlight a cardinality-minimal GIS for this formula that corresponds to the GICS in \cref{ex:small-network}.}
    \label{tab:truth-table}
\end{table}

\begin{example}
Let $\firevarset := \{\firevar_a, \ldots, \firevar_e\}$, $\sensorvarset := \{\sensorvar_a, \ldots, \sensorvar_e\}$, and $\groupset := \left\{\group_a := \{\firevar_a, \sensorvar_a\}, \ldots, \group_e := \{\firevar_e, \sensorvar_e\}\right\}$. 
Let $\maxidset=1$, and let $F_1(Z, \auxvarset)$ be defined as in \cref{eq:cnf}, with $Z = \support{\groupset}$.

After initialisation, $\unknown = \support{\groupset}$ and $\groupindsup = \varnothing$.
Let us assume that the algorithm now selects group $\group_e$ as the first group to test.
This causes both $\unknown$ and $\invariant$ to be updated to $\{\firevar_a, \sensorvar_a, \ldots, \firevar_d, \sensorvar_d\}$, and $\xi := \indvar_{x_a} \land \indvar_{y_a} \land \ldots \land \indvar_{x_d} \land \indvar_{y_d}$.
Let us assume that the algorithm first tests $\sensorvar_e \in \group_e$ for definability, constructing $\psi := \phi \land \xi \land \sensorvar_e \land \neg \hat{\sensorvar}_e$ and checking for satisfiability.
We inspect \cref{tab:truth-table} to check if $\psi$ is satisfiable.
As we can see in the table, there are no two rows that agree on the truth values of variables $\{\firevar_a, \sensorvar_a, \ldots, \firevar_d, \sensorvar_d\}$, but differ in their truth values of variable $\sensorvar_e$.
Hence, $\psi$ is unsatisfiable, and the algorithm moves to the second iteration of the inner `for'-loop to perform the same test for variable $\firevar_e$, finding again that $\psi$ is unsatisfiable.

The algorithm concludes that all variables in $\group_e$ are defined by the variables in $\invariant$, and moves on to test the next group.
Let us assume that the algorithm tests group $\group_d$ next.
It finds that $\group_d$ also does not belong in the GIS, and moves on to group $\group_c$.
Now we have $\unknown = \invariant := \{\firevar_a, \sensorvar_a, \firevar_b, \sensorvar_b\}$, $\xi := \indvar_{x_a} \land \indvar_{y_a} \land \indvar_{x_b} \land \indvar_{y_b}$, and $\psi := \phi \land \xi \land \sensorvar_c \land \neg \hat{\sensorvar}_c$.

Let us assume that the algorithm first checks $\sensorvar_c \in \group_c$.
Inspecting \cref{tab:truth-table}, we find that there are no two rows that agree on their truth values for $\firevar_a$, $\sensorvar_a$, $\firevar_b$ and $\sensorvar_b$, but disagree on their truth value for $\sensorvar_c$.
Hence, $\psi$ is unsatisfiable, and the algorithm moves on to test $\firevar_c$.

The rows $\{c\}$ and $\{e\}$ in \cref{tab:truth-table} agree on their truth values for $\firevar_a$, $\sensorvar_a$, $\firevar_b$ and $\sensorvar_b$, but disagree on their value for $\firevar_c$.
Consequently, $\psi$ is satisfiable, and we update $\groupindsup := \{\group_c\}$.

Let us assume that in the next iteration, the algorithm checks group $\group_b$.
It finds that, for both $\firevar_b$ and $\sensorvar_b$, $\psi$ is unsatisfiable, so $\group_b$ is discarded and not added to the GIS.
In the final iteration, we have $\unknown := \varnothing$, $\invariant := \{\firevar_c, \sensorvar_c\}$ and $\xi := \indvar_{x_c} \land \indvar_{y_c}$.
It is easy to see from \cref{tab:truth-table} that $\psi := \phi \land \xi \land \firevar_a \land \neg \hat{\firevar}_a$ is satisfiable (inspect rows $\{b\}$ and $\{e\}$), and thus the algorithm updates and returns $\groupindsup := \{\group_a, \group_c\}$.
\end{example}

\section{Experiments}
\label{sec:experiments}

In this section we describe our experiments aimed at evaluating the performance of \ouralgo{}, comparing it to the state-of-the-art ILP-based method.

\subsection{Experimental Setup}

\paragraph{Solving methods.}
We evaluate a method that encodes the problem into the CNF in \cref{eq:cnf} and then solves it by finding a GIS with \ouralgo{}.
In this section, we refer to this method as `\ouralgo{}'.
We compare the performance of \ouralgo{} to an ILP-based approach, as discussed in Section~\ref{sec:related-work}. We use {\ilpmethod}, based on initials of authors~\cite{PRP+20}, to refer to the ILP-based approach. 
We refer the reader to the extended version of this paper for details on implementation.

\paragraph{Software.}
Our implementation of \ouralgo{} uses SAT solver \cryptominisat{}~\cite{SNC09} version 5.11.7 (the latest version, last updated in December 2022) to determine the satisfiability of $\psi$ in \cref{line:is-independent} in \cref{alg:find-minimal-gis}.
We implemented the scripts for encoding networks into CNF (\cref{eq:cnf}) or ILP (\cref{sec:related-work}) with \python{}, using \pblib{}~\cite{PS15} for the CNF encoding of the cardinality constraint.
We solved the ILPs with \cplex{} 12.8.0.0.\footnote{Available at \url{www.ibm.com/analytics/cplex-optimizer}.}

\paragraph{Hardware.}
We ran our experiments on a high-performance cluster, where each node is equipped with two Intel E5-2690 v3 CPUs, each with 12 cores and 96~GB RAM, running at 2.60 GHz, under Red Hat Enterprise Linux Server 6.10.

\paragraph{Experimental parameters.}
We allow \ouralgo{} and \ilpmethod{} each one core, $3\,600$ CPU~s and $4$~GB RAM to encode and solve each (network, $\maxidset$) combination.
For \ouralgo{} we set a time limit of $\timelimit{} = 5\,000$ conflicts for the call to \cryptominisat{} in \cref{line:is-independent}.
For \cplex{} we use the default settings.
The running times we report are all user time measured in CPU~s.

\paragraph{Problem instances.}
Our benchmark set comprises $50$ undirected networks obtained from the {\em Network Repository}~\cite{RA15} and from the {\em IdentifyingCodes} \github{} repository~\cite{BS21a}.\footnote{Available at \url{https://networkrepository.com} and \url{https://github.com/kaustav-basu/IdentifyingCodes}.}, including grid-like networks, such as road networks and power networks, and social networks, such as collaboration networks and crime networks.
Their sizes vary from $10$ to $1\,087\,562$ nodes and $14$ to $1\,541\,514$ edges, and their median degrees vary from $1$ to $78$. 

\subsection{Research Questions}

The experiments in this section are aimed at answering the following main research questions:

\begin{description}
    \item[Q1] How many instances are solved by \ilpmethod{} and \ouralgo{}?
    \item[Q2] How do the solving times of \ilpmethod{} and \ouralgo{} scale with $\maxidset$ and $\abs{V}$?
    \item[Q3] How does the number of clause in the CNF encoding scale with $\maxidset$ and $\abs{V}$?
    \item[Q4] How do the cardinalities of the solutions returned by \ilpmethod{} and \ouralgo{} compare?
\end{description}

In summary, we find that \ouralgo{} solves nearly $10 \times$ more problem instances than \ilpmethod{} within the time limit of $3600$ seconds per problem instance.
The instances that can be encoded by both methods are solved up to $520 \times$ faster by \ouralgo{} than by \ilpmethod{}, depending on $\maxidset$.
On these instances, we find that the solution returned by \ouralgo{} is at most $60\%$ larger than that returned by \ilpmethod{}, but that most instances, the cardinality of the solution returned by \ouralgo{} is less than $10\%$ larger.
We find that the size of the CNF in \cref{eq:cnf} scales polynomially with $\abs{V}$ and $\maxidset$.
The largest problem that could be encoded and solved by \ilpmethod{} has $494$ nodes. 
The largest problem that could be encoded and solved by \ouralgo{} has $21\,363$ nodes; a $\sim 40$-fold improvement.

\subsection{Experimental Results}

In the remainder of this section, we describe our experimental results and answer our research questions.

\begin{table*}
    \centering
    \small
    \begin{tabular}{lrrrrrrrrrr}
\toprule
$\maxidset$         &               $1$ &              $2$ &              $3$ &              $4$ &              $6$ &              $8$ &             $10$ &             $12$ &             $16$ \\
\midrule
\ilpmethod{}   &   $5\,663$ ($11$) &   $6\,063$ ($8$) &   $6\,508$ ($5$) &   $6\,912$ ($2$) &   $6\,915$ ($2$) &   $6\,925$ ($2$) &   $6\,934$ ($2$) &   $6\,935$ ($2$) &   $6\,935$ ($2$) \\
\ouralgo{}   &      $969$ ($46$) &  $3\,146$ ($29$) &  $3\,346$ ($28$) &  $3\,425$ ($27$) &  $3\,097$ ($30$) &  $3\,062$ ($30$) &  $2\,959$ ($31$) &  $2\,940$ ($31$) &  $2\,393$ ($37$) \\
\bottomrule
    \end{tabular}
    \caption{PAR-2 scores and number of solved instances (in parentheses) for \ilpmethod{} and \ouralgo{}, and for each value of $\maxidset$ that we evaluated. The PAR-2 scores are given in CPU~s, and we used a timeout of $3600$~s. For each $k$, the total number of instances was $50$.
    }
    \label{tab:solved-instances}
\end{table*}

\paragraph{Q1: Number of solved instances.} 
We report the number of solved instances by \ouralgo{} and \ilpmethod{} for the $9$ tested values of $\maxidset$ in \cref{tab:solved-instances}. 
An instance is solved if the solving method terminates before the timeout time. In the case of \ilpmethod{}, this means that the returned solution is cardinality-minimal. In the case of \ouralgo{}, this means that the solution is (close to)\footnote{Because of the time limit $\tau$ in \cref{line:is-independent} of \cref{alg:find-minimal-gis}.} set-minimal.
Overall, \ouralgo{} solved $289$ of the $450$ problem instances, while \ilpmethod{} solved only $36$ out of $450$.
Hence, the \ouralgo{} solves over $8$ times as many problem instances as \ilpmethod{}. We now delve into the internals of {\ouralgo} and {\ilpmethod}:  \ouralgo{} was able to encode the GICS problem into CNF (\cref{eq:cnf}) for each value of $\maxidset$ for $49$ of the $50$ networks.
The largest network it could encode into CNF has $227\,320$ nodes and $814\,134$ edges.
We find that \ouralgo{} returned a GIS for most of these CNFs.
On the other hand, \ilpmethod{} was able to encode at most $11$ of the $50$ benchmarks into an ILP, which was for $\maxidset = 1$ (it performed worse as $\maxidset$ increased).
The largest network that it could encode has $494$ nodes and $1080$ edges.
For larger values of $\maxidset$, \ilpmethod{}'s ability to encode the networks drops rapidly, being only able to encode and solve $2$ out of the $50$ networks for $\maxidset \geq 4$.
These two networks are the smallest in our benchmark set, with only $10$ an $14$ nodes.

\begin{table*}
    \centering
    \small
    \begin{tabular}{lrrrrrrrrr}
    \toprule
    $k$ (\# instances)          &           $1$ ($11$) &                   $2$ ($8$) &           $3$ ($5$) &        $4$ ($2$) &           $6$ ($2$) &        $8$ ($2$) &       $10$ ($2$) &       $12$ ($2$) &       $16$ ($2$) \\ 
    \midrule
\ilpmethod{} &  $143.59$ &              $96.53$ &  $367.69$ &   $5.45$ &  $87.00$ &  $349.56$ &  $557.71$ &  $577.94$ &  $593.18$ \\
\ouralgo{} &    $4.01$ &  $5.11$ &    $1.31$ &  $1.14$ &  $1.17$ &   $1.16$ &     $1.10$ &     $1.15$ &              $1.14$ \\
    \bottomrule
    \end{tabular}
    \caption{Median solving times for the instances that could be encoded into ILP.}
    \label{tab:average-solving-times}
\end{table*}

\paragraph{Q2: Solving time.}
\Cref{tab:solved-instances} compares the PAR-2 scores\footnote{The PAR-2 score is a penalised average runtime. It assigns a runtime of two times the time limit for each benchmark the tool timed out on, or ran out of memory on.} of \ouralgo{} to those of \ilpmethod{}.
The \ouralgo{} method is up to $\sim 6$ times faster than \ilpmethod{}, in terms of PAR-2 scores, for smaller values of $k$ and $2 \times$ faster for larger values of $k$. 
Since {\ilpmethod} often times out during encoding phase (due to the blow-up of the size of the encoded formula), we also a provide comparison, in \cref{tab:average-solving-times}, for the instances for which the encoding phase of {\ilpmethod} did not time out and for which the underlying ILP solver did not timeout either, which was the case for all instances for which the encoding did not time out. 
It is worth remarking that all such instances were solved by \ouralgo{} as well.
Here, we find that \ouralgo{} is up to $\sfrac{593.18}{1.14} \approx 520\times$ faster than \ilpmethod{} in terms of median solving time.
Overall, our results that {\ouralgo} achieves significant performance improvements over {\ilpmethod}, in terms of running time.

\begin{figure}[t]
    \centering
    \includegraphics[width=\columnwidth]{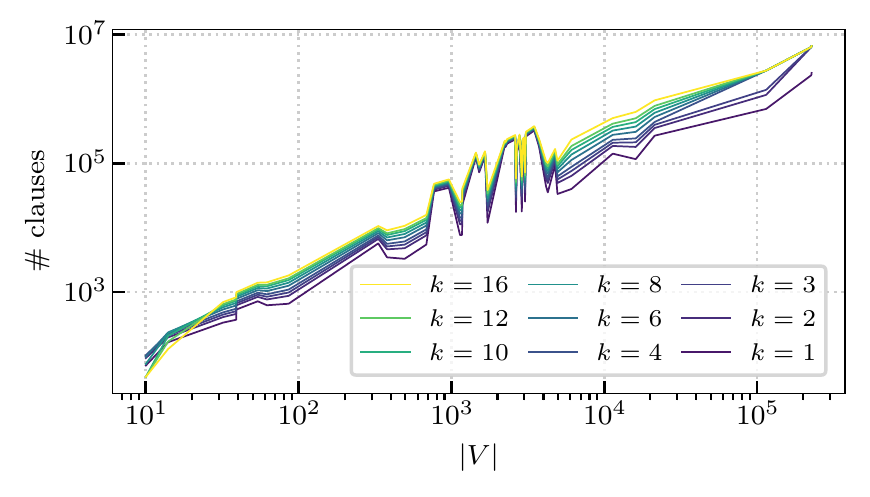}
    \caption{Number of clauses in the CNF encoding as a function of the number of nodes in the input network.}
    \label{fig:num_clauses_vs_num_nodes}
\end{figure}

\begin{figure}[t]
    \centering
    \includegraphics[width=\columnwidth]{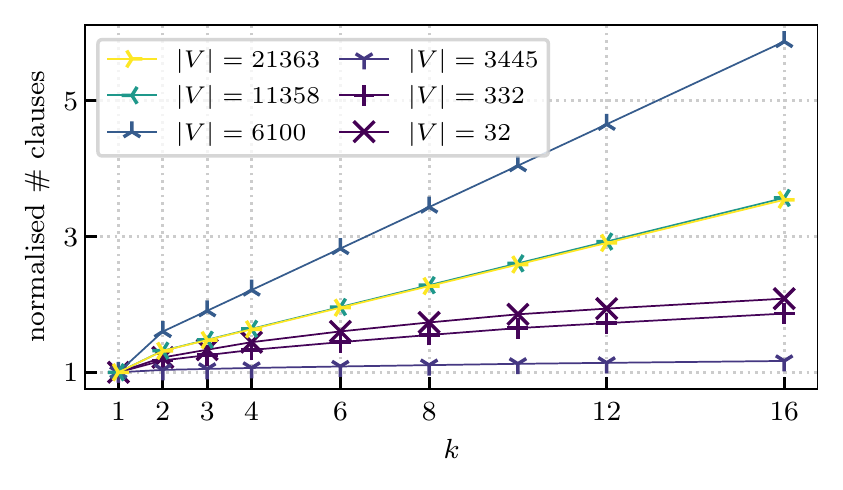}
    \caption{Normalised size of the CNF as a function of $\maxidset$, for selected networks. For each example, we divided the number of clauses in $F_{\maxidset}(Z, \auxvarset)$ by the number of clauses in $F_1(Z, \auxvarset)$, such that the model sizes are normalised \wrt{} the smallest model for each graph.}
    \label{fig:num_clauses_vs_k}
\end{figure}

\paragraph{Q3: Model size.}
\Cref{fig:num_clauses_vs_num_nodes} shows that the number of clauses in the CNF encoding scales polynomially with the number of variables in the instance. 
The oscillations in the plot can be explained by the $\abs{E}$-contribution to the CNF size (\cref{lem:num-clauses}). 
Our benchmark set contains both social networks (with high median degrees) and grid-like networks (with low median degrees), and hence with different densities.
\Cref{fig:num_clauses_vs_k} shows typical examples of how the number of clauses in the CNF encoding grows with increasing $\maxidset$.

\paragraph{Q4: Solution quality.}
We compared the quality of solutions returned by {\ilpmethod} and {\ouralgo} over the 36 instances that {\ilpmethod} could solve. 
In particular, we computed the ratio $\ratio := \sfrac{\abs{\groupindsup}}{\abs{\sensorset_{ILP}}}$, wherein $\groupindsup$ is the set computed by \ouralgo{}, while $\sensorset_{ILP}$ is the set computed by \ilpmethod{}.
In our experiments, we found that $1 \leq r \leq 1.6$, but for the majority of instances we found $\ratio < 1.1$. Furthermore, $4$ out of $36$ instances had a ratio $\ratio = 1$.
Hence, even in our naive implementation, the cardinalities of our solutions are almost as good as the minimum cardinality guaranteed by \ilpmethod{}.

\section{Conclusion}
\label{sec:conclusion}

In this paper, we focused on the problem of generalised identifying code set (GICS) problem which, given an input network, aims to find a set of nodes in which to place sensors in order to uniquely detect node failures, and where the number of placed sensors must be minimised.
We first identified the primary bottleneck of the prior state-of-the-art approach based on an ILP encoding: the blowup in the encoding size.
To address this shortcoming, we introduced {\em grouped independent support (GIS)} and reduced the GICS problem to the problem of finding a GIS of a Boolean formula.
Relying on the fact that algorithms for finding a minimised independent support are fast in practice, we designed and implemented an algorithm, \ouralgo{}, that finds a minimised, though not necessarily cardinality-minimal, {\em grouped} independent support.
Our empirical evaluation demonstrates that {\ouralgo} achieves significant performance improvements over the prior state-of-the-art technique in terms of running time, while producing solutions that tend to be close to cardinality-minimal. 

\section*{Acknowledgements}

This work was supported in part by National Research Foundation Singapore under its NRF Fellowship Programme [NRF-NRFFAI1-2019-0004], Ministry of Education Singapore Tier 2 grant MOE-T2EP20121-0011, and Ministry of Education Singapore Tier 1 Grant [R-252-000-B59-114].  The computational work for this article was performed on resources of the National Supercomputing Centre, Singapore \url{www.nscc.sg}.
One of the authors (Sen) acknowledges the support of the National University of Singapore during his sabbatical at NUS.
We thank Mate Soos for his help in debugging, Kaustav Basu for providing benchmark problems, and anonymous reviewers for their constructive feedback.

\FloatBarrier
\bibliographystyle{named}
\bibliography{ijcai23}

\end{document}